\documentclass[conference]{IEEEtran}
\IEEEoverridecommandlockouts
\usepackage{cite}
\usepackage{amsmath,amssymb,amsfonts}
\usepackage{algorithmic}
\usepackage{graphicx}
\usepackage{textcomp}
\usepackage{xcolor}
\def\BibTeX{{\rm B\kern-.05em{\sc i\kern-.025em b}\kern-.08em
    T\kern-.1667em\lower.7ex\hbox{E}\kern-.125emX}}
\begin{document}

\title{Identifying Difficult exercises in an  eTextbook Using Item Response Theory and Logged Data Analysis\\

}

\author{\IEEEauthorblockN{1\textsuperscript{st} Ahmed Abd Elrahman}
\IEEEauthorblockA{\textit{Information System Department} \\
\textit{Faculty of Computers and Information}\\
Assiut University\\
Egypt \\
ahmedabdo@aun.edu.eg}
\and
\IEEEauthorblockN{2\textsuperscript{nd} Ahmed Ibrahim Taloba}
\IEEEauthorblockA{\textit{Information System Department} \\
\textit{Faculty of Computers and Information}\\
Assiut University\\
Egypt \\
Taloba@aun.edu.eg}
\and
\IEEEauthorblockN{3\textsuperscript{rd} Mohammed F Farghally}
\IEEEauthorblockA{\textit{Information System Department} \\
\textit{Faculty of Computers and Information}\\
Assiut University\\
Egypt \\
mfseddik@aun.edu.eg}
\and
\IEEEauthorblockN{4\textsuperscript{th} Taysir Hassan A Soliman}
\IEEEauthorblockA{\textit{Information System Department} \\
\textit{Faculty of Computers and Information}\\
Assiut University\\
Egypt \\
taysirhs@aun.edu.eg}

}

\maketitle

\begin{abstract}
The growing dependence on eTextbooks and Massive Open Online Courses (MOOCs) has led to an increase in the amount of students' learning data. By carefully analyzing this data, educators can identify difficult exercises, and evaluate the quality of the exercises when teaching a particular topic. In this study, an analysis of log data from the semester  usage of the OpenDSA eTextbook was offered to identify the most difficult data structure course exercises and to evaluate the quality of the course exercises. Our study is based on analyzing students' responses to the course exercises. We applied item response theory (IRT) analysis and a latent trait mode (LTM) to identify the most difficult exercises .To evaluate the quality of the course exercises we applied IRT theory. Our findings showed that the exercises that related to algorithm analysis topics represented the most difficult exercises, and there existing six exercises were classified as poor exercises which could be  improved or need some attention.
\end{abstract}

\begin{IEEEkeywords}
interactive learning ,item response theory, eTextbooks, item characteristics curves, item Information Curve, data structures and algorithms.
\end{IEEEkeywords}

\section{Introduction}
The rising usage of interactive online course materials at all levels of education, including online eTextbooks, Massive Open Online Courses (MOOCs), and practice platforms like Khan Academy and Code Academy has spread especially with the spread of COVID-19 worldwide [1, 2, 3 and 4].In order to reduce its spread in educational institutions, most educational institutions have resorted to teaching their courses via online platforms.
During teaching of any online course, the interaction between students and educators takes place very little. It could be challenging for educators to know parts that students suffer from, as well as to assess the quality of exercises due to the lack of interaction with students. So knowing which topics students struggle with and attempting to improve or develop new methods to present these topics may be an essential step in enhancing the educational process and boosting the quality of the educational process. When students struggle with some issues in a course and no one strives to treat and simplify these topics. It is possible that they will drop out of the course and they will not finish studying the course resulting in failure in the educational process [2]. Knowing what topics students find difficult helps instructors to better allocate course resources. Based on the interactions of students with the OpenDSA eTextbook system [3, 5], we present techniques for automatically determining the most difficult exercises for students. The exercises students struggle with the most can be detected by experienced instructors, but this may frequently takes a long time and effort.
The topic of our study is a data structure and algorithms course (CS2). Our study has two aims, the first one is identifying the most difficult CS2 exercises, and the second is the evaluation of the quality of exercises in the CS2 course. To identify the most difficult exercises, we applied two different approaches, the first one is IRT theory and an LTM technique for analyzing student responses to exercises. LTM assumes that specific traits or characteristics can predict test performance [7]. IRT offers a model-based association between the test characteristics and the item responses [8]. The second approach involved analyzing how students interacted with exercises to see which ones were more challenging. We looked at how often students used hints. We found that the topics related to algorithm analysis have the most difficult exercises. To evaluate the course exercises, we also applied IRT. To classify each exercise as poor or good, we computed the item difficulty and item discrimination. Based on finding obtained, we found that six o exercises were classified as poor exercises that could be improved.   
In a sizable public institution, a CS2 course was taught using OpenDSA as the main eTextbook [1]. A module in OpenDSA represents one topic or portion of a typical lecture, such as one sorting algorithm, and is regarded as the most elementary functional unit for OpenDSA materials [2]. 
There is a range of various exercises in each module. One of these exercises requires the student to manipulate a data structure to demonstrate how an algorithm affects it. These are known as "Proficiency Exercises" (PE). PE exercises were developed and utilized for the first time in the TRAKLA2 system [9]. The other type of exercise is the Simple questions, which include various types of system questions such as true/false, multiple-choice, and short-answer questions. OpenDSA made utilized the exercise framework from Khan Academy (KA) [10] to save and present Simple questions.

 \section{Related Work}
 In [11], the responses of 372 students who registered in one first-year undergraduate course were utilized to evaluate the quality of 100 MCQs written by an instructor that was used in an undergraduate midterm and final exam. In order to compute item difficulty, discrimination, and chance properties they applied Classical test theory and IRT analysis models. The two-Parameter logistic (2PL) model consistently had the best fit to the data, they discovered. According to the analyses, higher education institutions need to guarantee that MCQs are evaluated before student grading decisions are made.
 In an introductory programming course, IRT was applied to assess students' coding ability [12]. They developed a 1PL Rasch model using the coding scores of the students. Their findings revealed that students with prior knowledge performed statistically much better than students with no prior knowledge.
In order to analyze the questions for the midterm exam for an introductory computer science course, the authors of [13] utilized IRT. The purpose of this study was to study questions’ item characteristic curves in order to enhance the assessment for future semesters.
The authors applied IRT for problem selection and recommendation in ITS. To automatically select problems, the authors created a model using a combination of collaborative filtering and IRT [14].

 \section{Experimental Analysis }
 Students make many interactions during their dealing with the eTextbook, every student interaction represents a log, and all student logs are stored in the OpenDSA system. OpenDSA contains different types of interactions. Interactions are divided into two types the first one is only interactions with the eTextbook itself, such as loading a page, reloading a page, clicking on a link to go somewhere else, or viewing slideshows. The second type is the interactions with all types of eTextbook exercises such that attempts to answer any exercise, submit an answer, or request a hint. This study focused more on the second type. In [2] a more description of interactions and exercise types.The amount of questions in each exercise varies.In this work, during the fall of 2020, we analyzed data of students who were enrolled in the  CS2 course. There are about 303,800 logs that represent the interactions of students with the eTextbook. These logs contain the name and description of the action, the time of the interactions, and which module the student executed the interactions on it.
As for the interactions of students with the exercises, we analysed about 200,000 logs .every log consists of time in which a student interacted with the question,total seconds in which a student finished interacting with a question,total count of hints that the student requested when interacting with a question,Total counts of attempts for student attempts to a question, and The type of request to a question (attempt or hint).In [15], different measures were applied in order to determine the difficult topics in a CS3 course. These measures are correct attempt ratio (r), difficulty level (dl), students' hint usage (hr), and incorrect answer ratio (it). We computed these measures for every exercise. In the next subsections, we will talk about them.

\subsection{Analysis of the ratios of right answers}
 We aim to give a value to every exercise in terms of “relative difficulty”. Our aim is to find which exercises average-ability students find comparatively difficult. From this, we intend to learn which themes are the most difficult for students. As a result, maybe lead us to refocus the instructional efforts. In the OpenDSA, students can answer an exercise as many times as they want until they get it correct. This will result in most students receiving almost full marks for their exercises[15]. Among the vulnerabilities, as is typical with online courseware that most students exploit is that some exercises can be "gamed" [16], In OpenDSA means that in order for students to get a question instance which easy to solve they reload the current page repeatedly. Due to the previous reasons, we have not counted the number of students who completed an exercise correctly. Instead, we employed other definitions for difficulty.  
To measure the exercise difficulty, we looked at the ratio of correct to incorrect answers in OpenDSA exercises, such that the correct attempt ratio for difficult exercises should be lower. We utilized the fraction r to evaluate student performance.
\begin{equation}\label{}
 r = \frac{\#\mspace{5mu}  of\mspace{5mu}  correct\mspace{5mu}  attempts}{\#\mspace{5mu}  of\mspace{5mu}  total\mspace{5mu}  attempts}      
          \end{equation}
          
We calculate the difficulty level (dl) for each exercise, such that 

\begin{equation}\label{}
 dl = \frac{1-\sum_{i}^{n} r(i)}{N}      
          \end{equation}
          
The number of students is referred to as {\bf N}, and the ratio of correct attempts is referred to as {\bf r}. 
In [15] ,the same measure was used to identify the difficult topics in CS3 course.In [17], similar measures was utilized to rate the difficulty of exercises, the authors utilized “the number of attempts it takes a student to figure out the right answer once making their initial mistake" as a metric of how difficult a logic exercises are. To determine the workout difficulty for an ITS, history of attempts conjunction with IRT was also applied in [18].
We categorized the exercises into categories based on their dl. The scores on the dl ranged from 0 to 0.85. The majority of the exercises in the fourth quartile ($dl > 0.34$) focused on algorithm analysis concepts (6 of a total of 11), one selection sort multiple-choice question, one recursion programming exercise, and one binary tree practice question.Exercises of the third quartile $(0.21\leq  dl\leq   0.34)$  covered mainly (7 of a total of 27) binary tree analysis. Six of these exercises covered linked list concepts. Four of these exercises covered sorting analysis, and three of these exercises covered the introduction chapter.The mechanics of algorithms or data structures were the focus of the exercises in the second quartile $(0.12 \leq  dl \leq  0.21)$ which covered   (20 out of 24). The remaining four were exercises that covered lists, queues, and an introduction chapter. The first quartile $(dl < 0.21)$  covered all exercises related to algorithms or data structure mechanics.The outcomes from previous results lead us to the conclusion that most of the difficult exercises belong to the fourth quartile that has the largest difficulty level values and this quartile contains most of the exercises that belong to the algorithm analysis concept. And the other quartiles that have less difficulty level value contain exercises related to the mechanism of algorithms and data structures, According to these findings, students had no trouble completing exercises that dealt with the behavior of algorithms and data structures. They seem to be having trouble with the analytical and algorithms exercises.
\subsection{The use of hints and Guessing
}
The measurement of "incorrect attempts" measure  that was used here not distinguish between utilizing a hint and giving an wrong answer. As a result, we took a closer look at the various types of wrong submissions for every exercise. We looked at how many hints were used in OpenDSA exercises and a trial-and-error technique was utilized to "guess" the answers. It is expected that more difficult exercises will show a higher hint rate and/or trial and error. To complete the exercise, the student must obtain a specific number correct (usually five) [6]. When a student submits an incorrect response, a point has been deducted from their credit toward this requirement. To avoid guessing, Students can also  use  more than one  hint to aid in understanding the answer to the question. The attempt in this situation is not assessed. 
The hint ratio (hr) was calculated for each exercise to analyze exercises based on students' hint usage \cite{r6}. 
 
\begin{equation}\label{}
hr = \frac{\#\mspace{5mu}  of\mspace{5mu}  hints\mspace{5mu}  used}{\#\mspace{5mu}  of\mspace{5mu}  hints\mspace{5mu}  used+\#\mspace{5mu}  of\mspace{5mu}  total\mspace{5mu}  attempts}      
          \end{equation}
          
We divided by (\# of   hints  used+\# of   attempts) ,Because it is possible that the number of hints is greater than the number of attempts, for example, it is possible that  when a student solves a specific exercise, he may have done two hints, but he attempts only once for this exercise. In this case, the hr value will become $> 0$.   

We computed the incorrect ratio ir for each exercise in order to analyze the exercises based on the rate of trial-and-error.

\begin{equation}\label{}
ir = \frac{\#\mspace{5mu}  of\mspace{5mu}  wrong\mspace{5mu}  answers}{\#\mspace{5mu}  of\mspace{5mu}  total\mspace{5mu}  attempts}      
          \end{equation}
 
{\bf Table 1}, contains eleven exercises with high percentages of hints or incorrect answers ratios. They related to topics covering Algorithm Analysis, Queues Analysis, Linked List and runtime analysis of bubble sort, and insertion ell sort. We observed that most exercises that had a low incorrect answer and a low hint ratio belong to Binary Trees arrays, introduction chapter, object-oriented programming, and lists. The reason for this is that most students are familiar with these concepts from previous courses. When used as a measure of exercise difficulty, a high rate of hint use is used, Algorithm analysis, Linked List Analysis, Queues Analysis, and Sorting Analysis exercises looked to be more "difficult" than others. The reason for this is that students may be not familiar with these concepts or they deal with these concepts for the first time in the CS2 course. 

\begin{table}[h]
    \centering
   \caption{ ir and hr for difficult exercises.}
    \label{data}
    \resizebox{0.5\textwidth}{!}{%
      \begin{tabular}{|c|c|c|c|}
   \hline
  {Exercise } & {hr}& {ir}& {Topic}  \\ \hline
 
     \cline{1-4} 
ListOverheadp & 0.66 & 0.6 2& ListOverheadAnalysis \\
   \cline{1-4} 
LqueueDequeuePROp & 0.40 & 0.61& Linked Dequeue Analysis  \\
       \cline{1-4} 
BubsortPROp & 0.37 & 0.33 & Bubble Sort Analysis \\
          \cline{1-4} 
AqueueEnqueuePROp & 0.35 & 0.73 & Array-Based Queue Enqueue\\
\cline{1-4} 
GrowthRatesPROp & 0.24 & 0.94 & Growth Rates \\
\cline{1-4} 
ListRemovePROp & 0.34 & 0.68 & Doubly Linked Lists\\
\cline{1-4} 
InssortPROp & 0.33 & 0.38 & Insertion Sort Analysis \\
\cline{1-4} 
AqueueDequeuePROp & 0.27 & 0.56 & Array-Based Queue Dequeue \\
\cline{1-4}  
AnlsIntroMCQtmcmp & 0.21 & 0.68 & Algorithm Analysis  \\  
 \cline{1-4}  
LLMCQchngcrsrp & 0.20 & 0.30 & Linear Structure Analysis  \\  
\cline{1-4}  
ComparingAlgorithmsSumm & 0.19 & 0.82 &Compared Algorithms  \\ 
  \hline
    \end{tabular}%
    } 
\end{table}

\subsection{IRT analysis }
IRT [15] examines item-level test behavior and provides feedback on the relative difficulty of different questions. According on the presumption that each response has a value of 0 or 1, many IRT models have been developed.We dichotomize the answers in order to perform IRT analysis. For  $ r \geq 0.70 $, we gave 1 point and for $r < 0.70$, we gave 0 point.every chapter was analyzed separately.In order to build a 2PL model for our study, we utilized R software specifically (ltm package) and built the 2PL model .

The equation for the  2PL model is given in equation below: -
\begin{equation}
f(x) = \frac {1}{1 + e^{-a(x-b)}}
\end{equation}
Where: e is the constant 2.718, {\bf b} stands for difficulty parameter,  {\bf a}  stands for discrimination parameter and x stands for ability level [19]. The logistic equation when graphed produces plots that are called item characteristic curves (ICC).We will discuss the two parameter (a and b) with brief details in the next sections.
For each OpenDSA exercise, we generated the Item Characteristic Curves (ICC), Item Information Curves (IIC) and Test Information Curves (TIF). Each curve's x-axis depicts the students' ability from -4 to 4. x = 0 denotes average ability. Given a student's ability, ICC shows the likelihood of a score of 1 and it allows us to characterize qualitatively whether these exercises are efficient or not. The IIC demonstrates how much information each exercise may inform us regarding the ability of students. The main purpose of the TIF is to determine the reliability of the test  at differentiating the different abilities of students. Students with above-average ability would be better distinguished by harder items. Easy items, on the other hand, would better differentiate students with below-average ability. The likelihood of earning a score of 1 for students with average ability may be shown on an ICC graph. Difficult items will obtain a $Pi (0) < 0.5$.

{\bf Introduction and Abstract Data Types exercises}: As illustrated in {\bf Fig 1}, almost all curves represent easy items because the likelihood of answering questions correctly for low-ability is high and approximately reaches 1 for high-ability examinees. So these exercises were previously familiar to students since they were covered in prerequisite courses. They helped us by providing information about students with below-average abilities$(x < 0)$.
\begin{figure}
\centering
\includegraphics[width=0.3\textwidth]{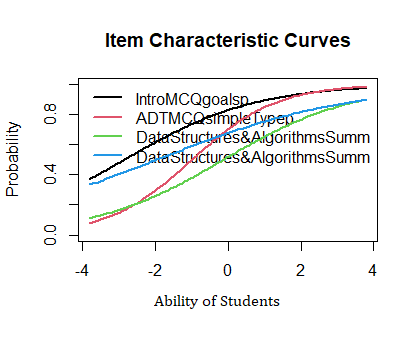}
\caption{\label{fig:pre}Abstract Data Types ICC.}
\end{figure} 
{\bf List Interface \&Array based Lists exercises}: As illustrated in {\bf Fig 2}, All curves represents easy items because the likelihood of answering questions correctly for low-ability is high and approximately reaches 1 for high-ability examinees. All Exercises in this chapter are easy and all students familiar with these exercises. Through these exercises, we were able to learn more about students with below-average abilities $(x < 0)$.
 \begin{figure}
\centering
\includegraphics[width=0.3\textwidth]{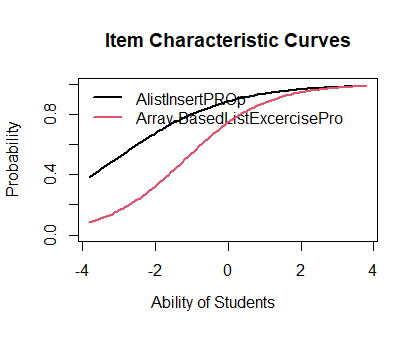}
\caption{\label{fig:pre}List Interface \&Array based Lists ICC.}
\end{figure} 
 
{\bf Algorithm analysis chapter exercises}: As illustrated in {\bf Figures 3 and 4}, almost all curves represent difficult items because the likelihood of answering questions correctly for most ability scales is low and increases only when reaching high-ability levels. Most students struggled with the exercises related to this chapter. As a result of these exercises, they helped us in giving information about students who have above-average abilities.

 \begin{figure}
\centering
\includegraphics[width=0.3\textwidth]{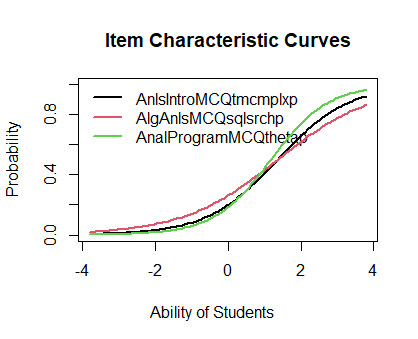}
\caption{\label{fig:pre}Algorithm Analysis ICC .}
\end{figure} 
 \begin{figure}
\centering
\includegraphics[width=0.3\textwidth]{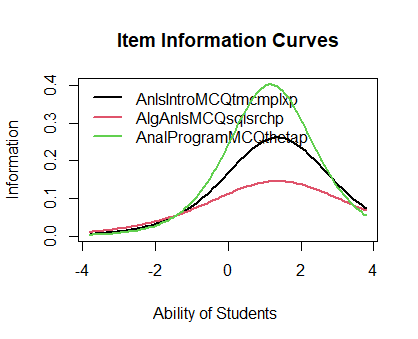}
\caption{\label{fig:pre}Algorithm Analysis IIC .}
\end{figure} 

{\bf Stacks exercises}: As illustrated in {\bf Fig 5}, most of the exercises in this chapter seem to be easy, and most of the students are familiar with them. As a result, Through these exercises, we were able to learn more about students with below-average abilities $(x < 0)$.

 \begin{figure}
\centering
\includegraphics[width=0.3\textwidth]{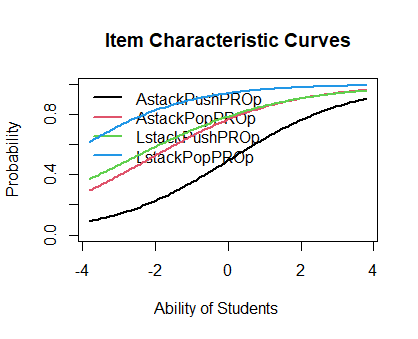}
\caption{\label{fig:pre}Stack ICC.}
\end{figure} 
{\bf Recursion exercises}: For students, just three exercises looked like they would be difficult. These involved forward flow tracing exercises, recursion programming exercises: subset-sum, and recursion programming exercises: Pascal triangle.These exercises helped us in providing information about students who have below-average abilities.
 {\bf Sorting exercises}: The chapter on sorting features the most exercises, all of which are of varying difficulty levels. More advanced sorting algorithms (Bubble sort, insertion sort, and selection sort) appeared to give more details about students with above-average ability $(x > 0)$.this chapter  exercises helped us in providing   information about students who have above-average abilities.
 
{\bf Linked List exercises}: The Linked List exercises have different difficulty levels. Exercises covering Linked List Remove, List element detection, and Linear Structure seemed  give more information about students with above-average ability $(x > 0)$. In general,these exercises  seem provided a good range of easy to tough exercises, as well as useful information for distinguishing between students of various abilities.
{\bf Binary Trees exercises}: The Binary Trees chapter has the most exercises. For students, only two exercises seemed to be difficult. Those two exercises cover preorder traversal and in order traversal.These exercises helped us in providing information about students who have below-average abilities.

The summary of previous results is that students struggle with almost all exercises related to the algorithm analysis chapter because the ICC and IIC curves for that chapter show that likelihood of answering questions correctly for most of ability-scale is low and increase only when reached to high-ability levels. Some exercises in Binary Tree, Linked List, Sorting,and Recursion seem to be difficult for students and they may need some attention. Almost most exercise-related to List and Abstract Data Type seems to be familiar to students.

We can conclude that by using the measures ir, hr, dl, and IRT analysis that the exercises belonging to the algorithm analysis module is the most difficult exercises in the CS2 course followed by some exercises in other modules   such that Binary Tree and Linked List, These modules may need some attention. While the exercises pertaining to the behaviour and mechanics of algorithms and data structures appeared to be familiar to the students.

\section{Evaluation of exercises' quality }
Such as we applied IRT analysis in determining the most difficult exercise, we applied it also in the Evaluation of exercises quality. We also applied   (2 PL) model and the test item analysis is based upon item discrimination (a) and item difficulty (b) .In [20] the (2PL) model was utilized to examine the quality of test items. We classified each exercise as poor or good exercise based upon its item discrimination and difficulty value. The next section describe the two terms. 
\subsection{(a) Parameter: Item discrimination}
One of the characteristics of a good test item is that it will be answered correctly by high-ability students more often than lower-ability. The (a) parameter reflects how effectively an item can differentiate between examinees of various abilities. A high discrimination level means the item can tell the difference between individuals who have high and low abilities. While most test items will have a positive value, some items may  have negative discrimination. In such items, the probability of correct response decreases because the ability level increases from low to high. This tells that something is wrong with the item and it's a warning that the item needs some attention. For many of the item response theory, the worth of the discrimination index value is positive [19]. 

\subsection{(b) Parameter: Item difficulty}
The point where the S-shaped curve has the steepest slope is the (b) parameter, which denotes an item's difficulty. The greater an examinee's ability level must be to successfully answer an item, the harder the item is. Items with high b values are challenging, meaning that low-ability test takers are unlikely to successfully respond. Values of b larger than 1 indicate a challenging item. Easy items are those with low b values below -1 [20].

\subsection{Results of exercises quality  }
As we said earlier, we built a (2PL) model to assess the quality of the exercises; we dichotomize the answers in order to perform IRT analysis. For $ r \geq 0.70 $, we gave 1 point and for $r < 0.70$, we gave 0 point, and we computed difficulty and discrimination index for each exercise and we used the two measures to classify the exercise as good or poor. 
The difficulty index and discrimination index for poor  exercises are shown in {\bf TABLE II}.
{\bf TABLE III} shows ranges of values used to describe an item’s discrimination level[19].

\begin{table}[h]
    \centering
   \caption{ Difficulty index and discrimination index for poor  exercises.}
    \label{data}
    \resizebox{0.5\textwidth}{!}{%
      \begin{tabular}{|c|c|c|c|}
   \hline
  {Exercise } & {Topic}& {Difficulty Index}& {Discrimination index}  \\ \hline
 
     \cline{1-4} 
AlistRemovePROp & Array-Based List (Remove) & 6.72-Hard& -0.4715-None \\
   \cline{1-4} 
CompareTF-MCQ5p & Sorting Terminology  & -2.24-Easy& 0.1614-Very Low \\
       \cline{1-4} 
SelSortPROp& Selection Sort  Analysis & -34.98-Easy & 0.0496-Very Low \\
          \cline{1-4} 
BTSummaryQuestionsp& Binary Tree Traversals & 2.20-Hard & -0.0303-None\\
\cline{1-4} 
BSTremovePRO&Binary Search Trees(Remove) & -0.20-Easy & 0.3297-Very Low \\
\cline{1-4} 
binarySearchPRO&Programming runtime & 8.02-Hard & -0.3379-None \\

  \hline
    \end{tabular}%
    } 
\end{table}
\begin{table}[h]
    \centering
   \caption{ Labels for item discrimination parameter value[19].}
    \label{data}
    \resizebox{0.25\textwidth}{!}{%
      \begin{tabular}{|c|c|}
   \hline
  {Label verbal } & {Range of values} \\ \hline
 
      \cline{1-2}  
None & 0   \\
 \cline{1-2}  
Very Low & 0.01-0.34  \\
\cline{1-2}  
Low & 0.35-0.64 \\
\cline{1-2}  
Low & 0.35-0.64 \\
\cline{1-2}  
Moderate & 0.65-1.34  \\
\cline{1-2}  
High & 1.35-1.69  \\
\cline{1-2}  
Very High& $>1.70$ \\   
  \hline
    \end{tabular}%
    } 
\end{table}
According to the results , according to the ranges in {\bf TABLE III}, there are three exercises have a negative discrimination value, and they  have a high difficulty value, this means that Supposed to students with high levels have an advantage in answering these exercises over students with low levels, but negative discrimination  value means that these exercises give preference to the student with a low abilities to solve these exercises over students with high ability levels ,this is a contradiction, so we classified these exercises as poor exercises.they  maybe need to be improved or maybe need to some attention. The reason for the improvement is that these exercises give preference to the student with a low level to solve this exercise about the student with a high level.There exist Three exercises that have a very low discrimination level and easy difficulty level, So we classified these three exercises on the basis that they are poor exercises because these exercises differentiate between students who have below-average abilities and, do not differentiate between different abilities levels students , so these exercises also maybe need to be improved or some attention. The poor exercises related to Binary Tree Traversals, Binary Search Tree, Selection Sort Analysis, Array-Based List, Sorting Terminology, and Algorithm Analysis topics.
\section{Conclusions and future work}
With the spread of COVID-19 worldwide, online eTextbooks have become more prevalent, when teaching a particular course via eTextbook; it has become necessary to identify the difficult exercises and to evaluate the quality for course exercises. This may allow instructors and instructional material creators to concentrate their efforts on the most difficult topics. Our study focused on CS2 course which was taught through eTextbook in a large a large public research institution during fall 2020. Our study was based on an analysis of students' responses to the CS2 course exercises.
Our study has two objective; the first is to identify the difficult exercises in the CS2 course,and the second to evaluate the quality of this course excercises .To identify the most difficult excercises ,we applied two approaches the first one is IRT and LTM to analyze the interactions of students with exercises. While the second one is analyzing how students interact with exercises to know which of them is more difficult than the other. We built a 2PL model for our analysis. Our findings showed that ,the exercises pertaining to the behaviour and mechanics of algorithms and data structures appeared to be familiar to the students,but they did face difficulty in dealing with the exercises related to algorithms analysis concepts. 
To evaluate the quality of the exercises; we applied IRT analysis and built a 2PL model, and we computed the difficulty and discrimination index for every exercise. We classified each exercise as poor or good based on these two metrics. The results showed that three exercises had negative discrimination values and were classifies as  poor exercises.we classified them as poor because , When solving these exercises, the student with a low ability is given preference over the student with high ability to solve these exercises. And there are three exercises that have low discrimination levels,they classified as poor  because these exercises don’t differentiate between students from different abilities levels . These poor exercise may need to some attention or improvements.The poor exercises covered topics including Binary Tree Traversals, Binary Search Tree, Selection Sort Analysis, Array-Based List, Sorting Terminology, and Algorithm Analysis topics. There are many interactions that each student makes with the OpenDSA eTextbook, a summary of interactions was explained in [2], so In the future. We will try to find the best sequence of the interactions so that the student gets the best benefit from the eTextbook.

\vspace{12pt}

\end{document}